\documentclass[pdflatex,sn-mathphys-num]{sn-jnl}

\usepackage[utf8]{inputenc} 
\usepackage{newunicodechar} 

\DeclareUnicodeCharacter{00E9}{\'e} 
\usepackage{graphicx}%
\usepackage{multirow}%
\usepackage{amsmath,amssymb,amsfonts}%
\usepackage{amsthm}%
\usepackage{mathrsfs}%
\usepackage[title]{appendix}%
\usepackage{xcolor}%
\usepackage{textcomp}%
\usepackage{manyfoot}%
\usepackage{booktabs}%
\usepackage{algorithm}%
\usepackage{algorithmicx}%
\usepackage{algpseudocode}%
\usepackage{listings}%

\begin{document}

\title[Article Title]{Potential Paradigm Shift in Hazard Risk Management: AI-Based Weather Forecast for Tropical Cyclone Hazards}


\author*[1]{\fnm{Kairui} \sur{Feng}}
\email{kelvinfkr2015@gmail.com}

\author[2]{\fnm{Dazhi} \sur{Xi}}
\author[3]{\fnm{Wei} \sur{Ma}}

\author[4]{\fnm{Cao} \sur{Wang}}

\author[5]{\fnm{Yuanlong} \sur{Li}}

\author[6]{\fnm{Xuanhong} \sur{Chen}}

\affil[1]{\orgdiv{The National Key Laboratory of Autonomous Intelligent Unmanned Systems}, \orgname{Tongji University}, \orgaddress{\city{Shanghai}, \postcode{200120}, \country{China}}}

\affil[2]{\orgdiv{} \orgname{Princeton University}, \orgaddress{\city{New Jersey}, \country{United States}}}

\affil[3]{\orgdiv{}\orgname{The Hong Kong Polytechnic University}, \orgaddress{\city{Hong Kong}, \country{China}}}

\affil[4]{\orgdiv{School of Civil, Mining, Environmental and Architectural Engineering}, \orgname{University of Wollongong}, \orgaddress{\city{Wollongong}, \state{NSW}, \postcode{2522}, \country{Australia}}}

\affil[5]{\orgdiv{School of Atmospheric Sciences}, \orgname{Nanjing University}, \orgaddress{\city{Nanjing}, \country{China}}}

\affil[6]{\orgdiv{USC-SJTU Institute of Cultural and Creative Industry}, \orgname{Shanghai Jiao Tong University}, \orgaddress{\city{Shanghai}, \country{China}}}


\abstract{The advents of Artificial Intelligence (AI)-driven models marks a paradigm shift in risk management strategies for meteorological hazards. This study specifically employs tropical cyclones (TCs) as a focal example. We engineer a perturbation-based method to produce ensemble forecasts using the advanced Pangu AI weather model. Unlike traditional approaches that often generate fewer than 20 scenarios from Weather Research and Forecasting (WRF) simulations for one event, our method facilitates the rapid nature of AI-driven model to create thousands of scenarios. We offer open-source access to our model and evaluate its effectiveness through retrospective case studies of significant TC events: Hurricane Irma (2017), Typhoon Mangkhut (2018), and TC Debbie (2017), affecting regions across North America, East Asia, and Australia. Our findings indicate that the AI-generated ensemble forecasts align closely with the European Centre for Medium-Range Weather Forecasts (ECMWF) ensemble predictions up to seven days prior to landfall. This approach could substantially enhance the effectiveness of weather forecast-driven risk analysis and management, providing unprecedented operational speed, user-friendliness, and global applicability.}




\maketitle
\section{Introduction}
The integration of AI in weather forecasting is transforming various sectors, especially in the management of risks associated with meteorological hazards \cite{lagerquist2017machine,chen2022rainnet}. This paper focuses on risk management prior to tropical cyclones (TCs). Accurate and timely forecasts are crucial for making critical decisions related to emergency resource distribution, rescue operations, public alerts, and evacuation plans \cite{davidson2006hurricane, nateghi2011tropical, roy2012tropical,feng2022tropical,feng2022modeling}. Earlier knowledge of TC trajectories allows for more effective preparations, such as pre-allocating resources for power systems to expedite recovery post-TC, or issuing evacuation orders with greater precision for residents to safely relocate \cite{regnier2008public, murray2014assessment, kantha2013forecasting}. Although numerical weather prediction (NWP) models such as the Weather Research and Forecasting (WRF) model produce advanced and physically-based predictions for future meteorological hazards, their substantial computational requirements often limit the exploration of multiple scenarios. \cite{Skamarock2008,Bauer2015,bauer2021digital}.

An ensemble forecast in NWP generates multiple predictions using varied initial conditions or model parameters to address the uncertainty inherent in weather forecasting \cite{gneiting2005weather}. This technique enhances the reliability of forecasts, improves decision-making for weather-sensitive activities, and helps manage risks associated with severe weather by providing a range of possible outcomes. Despite its benefits, ensemble forecasting requires extensive computational resources, as each forecast iteration requires multiple model runs, leading to significant data process and storage needs, and increased operational costs.

Our research introduces a perturbation-based approach for generating ensemble weather forecasts for Pangu AI weather model \cite{bi2023accurate}. The AI-driven model rapidly produces thousands of unique ensemble scenarios with ease of use, without the need of fine-tune with expert knowledge of meteorology, making it accessible to a broader range of decision-makers \cite{Leutbecher2008,gneiting2014probabilistic}.By fine-tuning the extent of the perturbation, we adjust the ensemble prediction based on the Pangu model to achieve an uncertainty level comparable to that of the European Centre for Medium-Range Weather Forecasts (ECMWF) model.

This study analyzes three historical TCs: hurricanes—Irma (2017), Typhoon Mangkhut (2018), and TC Debbie (2017) — representing diverse contexts across three continents. The comparative analysis with the ECMWF ensemble forecasts \cite{Haiden2018} shows that our AI-generated predictions maintain similar accuracy and spatial patterns up to a week before landfall.

We have made our code publicly available to enable the community to replicate and expand upon our work. The code facilitates tasks such as dataset preparation, weather forecasting, and wind field visualization, and can be directly executed using Colab in a web browser-based environment.

We believe AI-driven weather forecasts unlock the potential for more comprehensive risk analyses, augmenting the capacity of disaster management agencies to mitigate TC impacts \cite{Knutson2019}. The global scope of our study demonstrates the model's adaptability and scalability, offering an applicable solution.

\section{Summarizing Recent AI-driven weather forecast models}

In recent years, AI-driven weather forecast models have rapidly evolved. We summarize several open-source AI-driven weather forecast models in Table~\ref{tab:my_label}. These models are typically trained on the ECMWF reanalysis dataset (ERA5) \cite{hersbach2020era5}, which offers a resolution of $0.25^{\circ}$ (on a $721\times 1440$ lattice) and is globally recognized for effective medium-range weather forecasting, with similar accuracy as current Integrated Forecast System(IFS) models. ERA5 contains 137 variables across various pressure levels; however, these AI-driven models often utilize only 10-40 selected variables for one time step to forecast the same variables for the next time step (for a fixed time interval, for example, one day). This approach enables flexible future projections through auto-regression. The open-source models are efficiently packaged, allowing predictions to be executed with a single line of code once the input data matrix is prepared. AI-driven models typically lack randomness or ensemble projections for the future due to the deterministic nature of neural network strictures, with some models based on Convolutional neural networks (CNNs) and others on transformers. Consequently, these models might not be directly suitable for risk management, as the deterministic outcome produced by an AI-driven model cannot be assumed to represent the ground truth. In response, this paper explores a perturbation-based method to generate ensemble forecasts using these AI-driven weather forecast models, enhancing their ability in risk management scenarios.

The primary advantage of AI models lies in their rapid inference time, which is the time required for a machine learning model to simulate a weather step given initial conditions on specific hardware. The main challenge in utilizing AI-driven weather forecast models is their high demand for GPU memory, often exceeding the capacity of consumer-grade GPUs like the 4090(24GB,\$1,500-\$2,000); typically, a V100 GPU (32GB,\$8,000-\$10,000) is necessary for efficient inference. For instance, inferring the Pangu model on a CPU takes over 7 minutes per step, whereas it takes less than one second on an A100 GPU(40GB-80GB,\$10,000-\$20,000).

In this project, we provide open-source code (\url{https://github.com/kelvinfkr/Perturbation_AI_weather}) hosted on the Colab platform, a cost-effective online Python execution environment offering access to an A100 GPU for approximately \$10(USD) per 20 hours, as of April 2024. This \$10 enables more than 72,000 days, or over 20 years, of weather simulation, making AI-driven model a highly economical tool for risk analysis. The code is web-compilable, allowing users to interact with the model at minimal cost. The necessary ERA5 data can be downloaded from the official website or through a Google storage image integrated into our code, which supports downloading 10 GB of data in just 10 seconds. Although the full ERA5 dataset is extensive ($\sim 5 PB$), the data required to simulate one day's weather is less than 10 GB, which our code can manage efficiently.

Recent studies have begun to cross-compare AI-driven weather forecast models, such as Weatherbench 2 \cite{rasp2023weatherbench} project which compares the prediction quality between multiple AI-driven models. Although Pangu is a leading model as of 2023, it may not be the most advanced in all respects. Users may select models based on specific needs; for example, Pangu does not offer precipitation data, whereas GraphCast does. Selection may also be based on computational efficiency — among these models, FuXi requires the least computational resource; or on the GPU model available - for those short for GPU resources, SphericalCNN might be preferable due to its lower GPU memory usage.

This paper employs Pangu as an illustrative example; however, the framework is generalizable to any AI-driven model.

\begin{table}[htbp]
    \centering
    \caption{List of Recent Open Source AI-Driven Weather Forecast Models}
{\tiny\begin{tabular}{c|c|c|c|c|c|c|c}
\hline Model/ Dataset & Type & \begin{tabular}{l} 
Initial con- \\
ditions
\end{tabular} & $\Delta x$ & Levels & \begin{tabular}{l} 
Training \\
data
\end{tabular} & \begin{tabular}{l} 
Training re- \\
sources
\end{tabular} & \begin{tabular}{l} 
Inference \\
time
\end{tabular} \\

\hline Pangu-Weather\cite{bi2023accurate} & Forecast & ERA5 & $0.25^{\circ}$ & 13 & \begin{tabular}{l} 
ERA5 (1979- \\
$2017)$
\end{tabular} & \begin{tabular}{l}
16 days; 192 \\
V100 GPUs
\end{tabular} & \begin{tabular}{l} 
several sec- \\
onds; single \\
GPU
\end{tabular} \\
GraphCast\cite{lam2022graphcast} & Forecast & ERA5 & $0.25^{\circ}$ & 37 & \begin{tabular}{l} 
ERA5 (1979- \\
$2019)$
\end{tabular} & \begin{tabular}{l}
4 weeks; 32 \\
TPU v4
\end{tabular} & \begin{tabular}{l}
$\sim 1 \quad$ minute; \\
single TPU
\end{tabular} \\
FuXi\cite{chen2023fuxi} & Forecast & ERA5 & $0.25^{\circ}$ & 13 & \begin{tabular}{l} 
ERA5 (1979- \\
$2017)$
\end{tabular} & \begin{tabular}{l}
$\sim 8$ days; 8 \\
A100 GPUs
\end{tabular} & \begin{tabular}{l}several seconds;\\ single
GPU\end{tabular}\\
SphericalCNN\cite{esteves2023scaling} & Forecast & ERA5 & $1.4 \times 0.7^{\circ}$ & 7 & \begin{tabular}{l} 
ERA5 (1979- \\
$2017)$
\end{tabular} & \begin{tabular}{l}
4.5 days; 16 \\
TPU v4
\end{tabular} &\begin{tabular}{l}$\sim 1 \quad$ minute; \\
single TPU\end{tabular}\\

FourCastNet\cite{pathak2022fourcastnet} & Forecast & ERA5 & $0.25^{\circ}$ & 20 & \begin{tabular}{l} 
ERA5 (1979- \\
$2017)$
\end{tabular} & \begin{tabular}{l}
16 hours; 64 \\
A100 GPUs
\end{tabular} & \begin{tabular}{l}several seconds;\\ single
GPU\end{tabular}\\

\hline
\end{tabular}}
    
    \label{tab:my_label}
\end{table}

\section{Weather Forecast}
\begin{figure}[ht]
\centering
\includegraphics[width=0.9\textwidth]{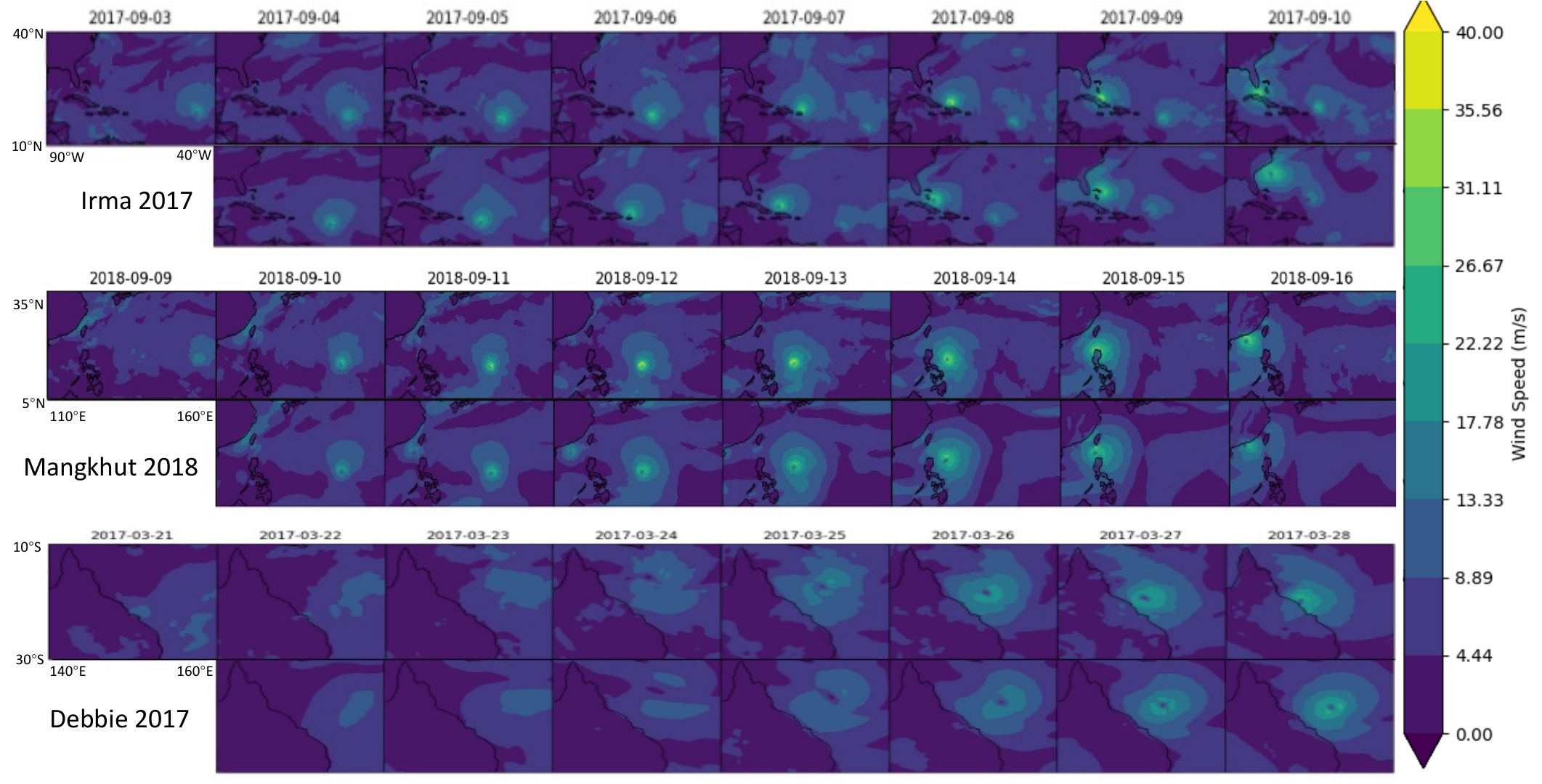}
\caption{Comparative Weather Forecasts by Pangu 7 Days Before Landfall: Truth surface Wind speed (ERA5 reanalysis \cite{hersbach2020era5}; upper panel) vs. Pangu's predictions (lower panel) for a) Hurricane Irma (2017), b) Typhoon Mangkhut (2018), and c) TC Debbie (2017).}
\label{fig:weather}
\end{figure}

In this paper, we highlight the Pangu AI model as an example to show the potential of AI-driven models in pre-hazard risk management. We employed the Pangu model with ERA5 reanalysis data to forecast surface wind fields of TCs seven days before landfall, following the methodology outlined in the Pangu paper \cite{bi2023accurate}. As illustrated in Fig.~\ref{fig:weather}, this AI-driven approach not only accurately captures the trajectories and intensification processes of TCs but also exhibits an exceptional ability to obtain the asymmetrical features of wind fields — a significant improvement over traditional wind field models. Each simulation was completed within seconds, demonstrating the feasibility of real-time ensemble forecasting. However, the model occasionally mispredicts detailed landfall locations, such as forecasting Irma's landfall on the eastern side of Florida instead of the western side. This fact presents a critical challenge in pre-TC risk management, where accurate landfall predictions are crucial for issuing effective evacuation orders, and highlights the need to integrate ensemble methods into the AI-driven model to enhance its reliability in managing such uncertainties.

\section{Perturbation based Ensemble}
In this study, we employed an engineered approach to generate ensemble weather forecasts using the Pangu model. The ensembles were created through systematic perturbation of the initial inputs into the Pangu model. The initial inputs have a structure consisting of 37 variables on each of the $721\times 1440$ lattice points for a specific time point. The output is the forecasted weather condition one day later, also with a $37\times 721\times 1440$ structure. By introducing stochastic variations -- specifically, shifting each input variable by 0, 1, 2, ..., up to $n$ hours from the ERA5 dataset -- we achieved a level of uncertainty quantification analogous to that in ECMWF's ensemble forecasts. Intuitively, and as demonstrated in the experiment, as the value of $n$ increases, the level of uncertainty in the model's predictions also rises.

We tuned the parameter $n$ and found that when $n=3$, the uncertainty in the Pangu ensemble matched that of current NWP models. The following analysis was conducted with $n=3$ as the perturbation level. This perturbation-based approach may not only be useful to create ensemble, but could also be useful for benchmarking different weather forecast models, as a robust model should maintain accuracy against such perturbations \cite{brenowitz2024practical}. While this perturbation approach is not strictly physical and cannot guarantee the prediction of a physically consistent environmental field for future weather, it provides a straightforward method to demonstrate the potential of AI-driven models to generate ensemble forecasts for meteorological hazards. In the future, this simple perturbation method is likely to be replaced by more sophisticated ensemble frameworks that offer greater physical accuracy. 

\begin{figure}[ht]
\centering
\includegraphics[width=0.9\textwidth]{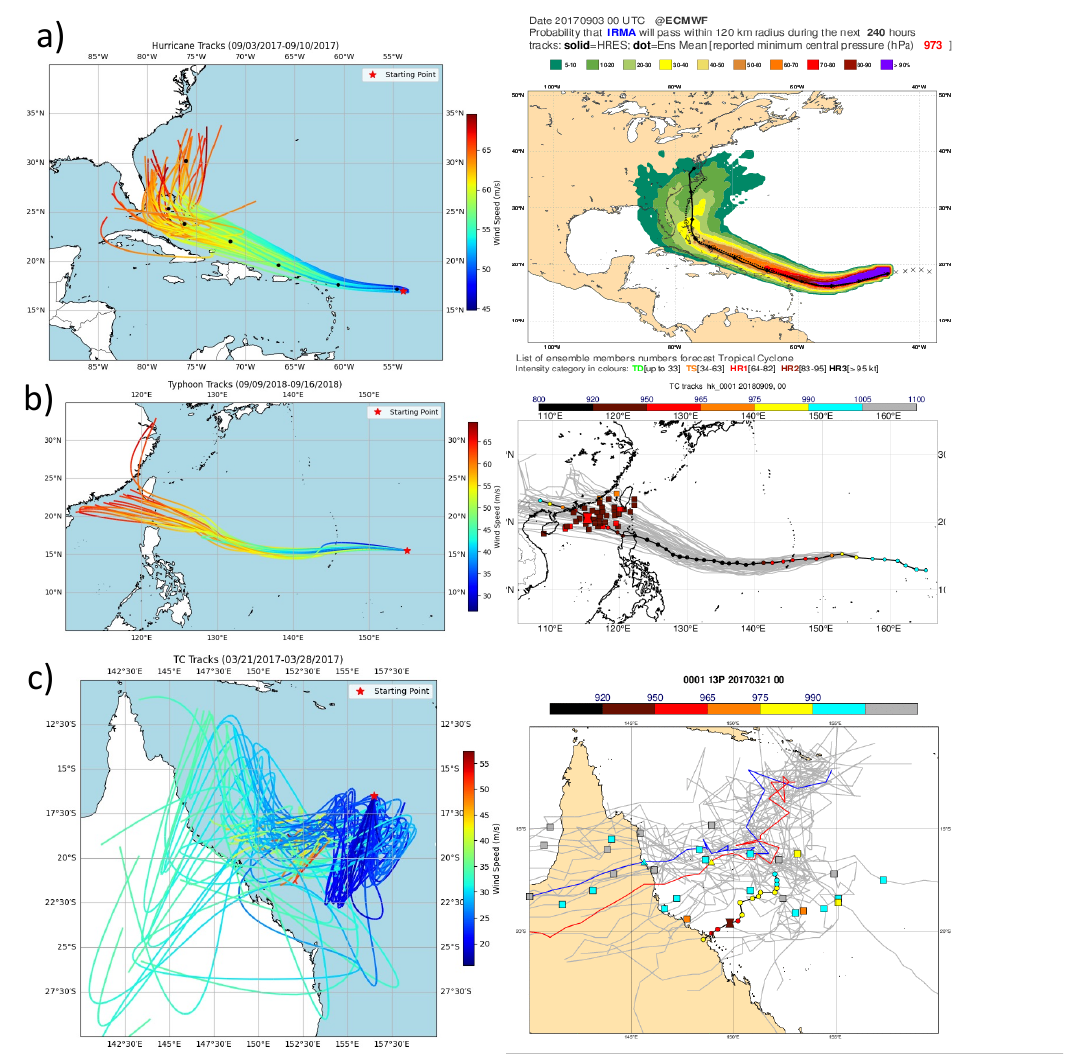}
\caption{Comparison of Ensemble Weather Forecasts: TC trajectories under Pangu Perturbations (left panel) vs. ECMWF Ensemble  \cite{Haiden2018} (right panel) for a) Hurricane Irma (2017), b) Typhoon Mangkhut (2018), and c) TC Debbie (2017).}
\label{fig:ensemble}
\end{figure}

As shown in Fig.~\ref{fig:ensemble}, the perturbation-based ensembles consistently yield realistic forecasts of the trajectory and intensity fluctuations of cyclones. For example, the model accurately predicted the strengthening of Hurricane Irma under the influence of the westerlies jet stream if it were to track eastward across the Gulf Coast. In contrast, a westward trajectory would result in a weakening effect. This understanding of meteorological dynamics confirms the Pangu model's comparability with established NWP models. Notably, in cases where NWP models perform well, such as Typhoon Mangkhut, the Pangu model also shows good performance. Conversely, in situations with high uncertainty, such as TC Debbie, where NWPs cannot conclusively predict outcomes, the Pangu model similarly reflects the high level of uncertainty. 

More in-depth and numerically, we computed the projected trajectory uncertainty between perturbation-based AI-driven ensemble and the ECMWF ensemble. For Irma 2017, the 7-day trajectory probability difference, quantified by the root mean square error (RMSE), between Pangu and ECMWF within a $5^\circ \times 5^\circ$ grid is $5.8\%$. The differences are $2.7\%$ and  $13.9\%$ for Mangkhut 2018 and Debbie 2017, respectively. These results demonstrate the consistency in uncertainty levels and trajectory projections between the proposed model and ECMWF ensembles. Furthermore, these results highlight the strong potential of substituting ECMWF models with the proposed model in risk management problems that require uncertain TC forecasts.

\section{Potential Applications and Further Adjustment}

\begin{figure}[ht]
\centering
\includegraphics[width=0.9\textwidth]{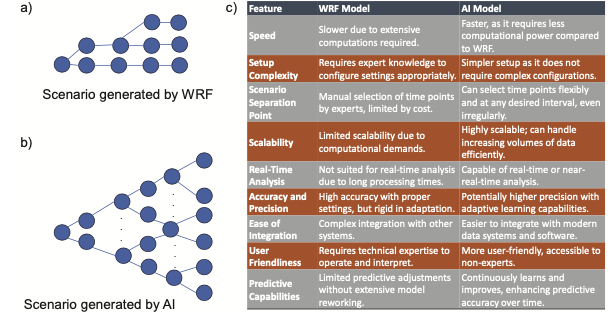}
\caption{Comparative Analysis of Meteorological Hazard Simulation Scenarios in Decision-Making: WRF vs. AI Models. (a) Limited scenario simulation with WRF: Ensembles triggered by expert judgement. (b) Extensive decision tree simulations using AI: Probabilistic and high-branching. (c) Illustrative comparison between WRF model and AI-driven models.}

\label{fig:framework}
\end{figure}

In a traditional framework for pre-meteorological hazard risk management, as shown in Fig.~\ref{fig:framework}a, decision-makers first obtain a scenario tree from weather forecast agencies. Typically, each scenario progresses directly to the end of the decision-making cycle without branches. A new branch of simulation for a specific scenario is only initiated when considered necessary by experts. Given the real-time requirements for pre-TC risk management, the number of scenarios considered in decision-making is generally fewer than 20.

Conversely, in a prospective AI-driven framework shown in Fig.~\ref{fig:framework}b, decision-makers can easily generate thousands of scenarios, each characterized by a Markovian relationship. 
Hence, modern optimization frameworks such as optimal control, Markovian decision process, smart prediction and optimization, and reinforcement learning could be introduced to enhance risk management. In contrast, in the traditional framework, due to the large uncertainty in weather forecasts, fuzzy math-driven optimizations, such as robust optimization, play more crucial roles.

Figure~\ref{fig:framework}c compares the traditional WRF Model with an AI-driven weather model across various key features relevant to meteorological applications. The WRF Model, while highly accurate when properly configured, is notably slower due to its intensive computational demands and requires significant expertise and manual effort for setup and integration with other systems. Its limited scalability and inappropriateness for real-time analysis hinder its use in emergency meteorological situations.

In contrast, the AI model has faster processing speeds as it exploits modern computational architectures and advances in parallel computing, making it ideal for real-time or near-real-time analysis. Its setup is simpler, escaping the need for complex configurations, and it exhibits high scalability, capable of handling large volumes of data efficiently. This model can select time points for scenario analysis flexibly, even at irregular intervals, without the cost constraints associated with the WRF Model. Moreover, the AI Model's ability to continuously learn and adapt enhances its predictive accuracy over time, allowing for more precise risk management under changing conditions. It is also more user-friendly and easier to integrate with modern data systems and software, making it more accessible to non-experts.

Therefore, it is foreseeable that AI-based weather forecasting, which is worth significant attention, could transform the landscape and paradigm of pre-meteorological hazard risk management.

\section{Discussion and Conclusion}
As an emerging technology, though performs well, the reliability of AI-driven weather forecast models has not yet been thoroughly proven. To better understand their dependability, these models should undergo further validation and testing.The significant potential of these models underscores the necessity for ongoing research aimed at evaluating their efficacy within comprehensive risk management frameworks. Future advancements in AI weather modeling should prioritize the integration of a wider array of risk factors, including precipitation, which is important due to its direct implications for economic impact and flood potential.

Moreover, utilizing AI-driven weather forecasting tools enhances the timeliness and accuracy of information, which is essential for organizing effective emergency responses. This capability could significantly support public safety and optimize resource allocation during large meteorological hazard events, thereby augmenting the resilience of affected communities and infrastructures.

\section*{Acknowledgements}
Kairui Feng thanks the support by National Natural Science Foundation of China (Grant No. 62088101).

Cao Wang was supported by the Australian Government through the Australian Research Council's Discovery Early Career Researcher Award (DE240100207).

Wei Ma was supported by the Research Institute for Sustainable Urban Development (RISUD) at the Hong Kong Polytechnic University (Project No. P0038288).

\section*{Declaration}
We have no conflicts of interest to declare.




\bibliography{sn-bibliography}

\end{document}